\documentclass[aps,showpacs,nofootinbib]{revtex4}

\usepackage{graphicx}
\usepackage{graphics}
\usepackage{amssymb}
\usepackage{amsmath}
\usepackage{bm}
\newcommand{\half}{\textstyle{\frac{1}{2}}}
\newcommand{\sT}{{\scriptscriptstyle T}}
\newcommand{\g}{\gamma}

\newcommand{\be}{\begin{equation}}
\newcommand{\ee}{\end{equation}}
\newcommand{\bea}{\begin{eqnarray}}
\newcommand{\eea}{\end{eqnarray}}
\newcommand{\beal}{\begin{align}}
\newcommand{\eal}{\end{align}}
\newcommand{\bespl}{\begin{split}}
\newcommand{\espl}{\end{split}}
\newcommand{\nn}{\nonumber}
\newcommand{\nslash}{\kern 0.2 em n\kern -0.50em /}
\newcommand{\kslash}{\kern 0.2 em k\kern -0.45em /}
\newcommand{\pslash}{\kern 0.2 em p\kern -0.50em /}
\newcommand{\Sslash}{\kern 0.2 em S\kern -0.50em /}
\newcommand{\Pslash}{\kern 0.2 em P\kern -0.50em /}
\newcommand{\Rslash}{\kern 0.2 em R\kern -0.50em /}
\begin{document}
\title{
Monte Carlo simulation of the Sivers effect in high-energy proton-proton 
collisions}

\author{A.~Bianconi}
\email{andrea.bianconi@bs.infn.it}
\affiliation{Dipartimento di Chimica e Fisica per l'Ingegneria e per i 
Materiali, Universit\`a di Brescia, I-25123 Brescia, Italy, and\\
Istituto Nazionale di Fisica Nucleare, Sezione di Pavia, I-27100 Pavia, Italy}

\author{Marco Radici}
\email{marco.radici@pv.infn.it}
\affiliation{Dipartimento di Fisica Nucleare e Teorica, Universit\`{a} di 
Pavia, and\\
Istituto Nazionale di Fisica Nucleare, Sezione di Pavia, I-27100 Pavia, Italy}

%%%%%%%%%%%%%%%%%%%%%%%%%%%%%%%%%%%%%%%%%%%%%%%%%%%%%%%%%%%%
\begin{abstract}
We present Monte Carlo simulations of the Sivers effect in the polarized Drell-Yan 
$pp^\uparrow \to \mu^+ \mu^- X$ process at the center-of-mass energy $\sqrt{s}=200$ 
GeV reachable at the Relativistic Heavy-Ion Collider (RHIC) of BNL. We use two 
different parametrizations for the Sivers function, one deduced from the analysis of
Semi-Inclusive Deep-Inelastic Scattering (SIDIS) data at much lower energies, and
another one constrained by the RHIC data for the $pp^\uparrow \to \pi X$ process at
the same energy. For a given luminosity of $10^{32}$ cm$^{-2}$ sec$^{-1}$, we
explore the necessary conditions to reach a statistical accuracy that allows to
extract unambiguous information on the structure of the Sivers function. In
particular, we consider the feasibility of the test on its predicted universality
property of changing sign when switching from SIDIS to Drell-Yan processes.
\end{abstract}


\maketitle

%%%%%%%%%%%%%%%%%%%%%%%%%%%%%%%%%%%%%%%%%%%%%%%%%%%%%%%%%%%%
\section{Introduction}
\label{sec:intro}

Azimuthal asymmetries in hard collisions involving (polarized) hadrons represent
a formidable testground for Quantum ChromoDynamics (QCD) in the nonperturbative 
regime. Since almost thirty years, data have been collected for hadron-hadron
collisions, and recently also for semi-inclusive $\g^\ast$-hadron processes 
in the regime of Deep-Inelastic Scattering (DIS). Large asymmetries were 
observed in the azimuthal distribution of final-state products (with respect to
the normal to the production plane), particularly when flipping the transverse
polarization of one hadron involved in the initial or final state: the socalled
Single-Spin Asymmetries (SSA). Examples of such SSA are found for the $pp\to
\Lambda^\uparrow X$ process~\cite{Bunce:1976yb} at forward rapidity, for the 
$pp^\uparrow \to \pi X$ 
process~\cite{Adams:1991cs,Adams:2003fx,Adler:2005in,Videbaek:2005fm} again at 
forward rapidity, and for the semi-inclusive hadron production $\g^\ast 
A^\uparrow \to \pi X$, where $A$ is the 
proton~\cite{Bravar:2000ti,Airapetian:2004tw,Diefenthaler:2005gx,Avakian:2005ps} 
or the deuteron~\cite{Alexakhin:2005iw}. Apart from the last one, in all other
cases SSA up to 40\% were detected which were totally unexpected, since they 
cannot be easily accommodated in a consistent manner within the perturbative QCD 
in the collinear massless approximation~\cite{Kane:1978nd}; moreover, they seem 
to persist also at higher energies typical of the collider 
regime~\cite{Adams:2003fx,Adler:2005in,Videbaek:2005fm}, which also contradicts 
QCD expectations. The same conclusion holds also for unpolarized Drell-Yan 
experiments at high energy like $\pi A \to \mu^+ \mu^- 
X$~\cite{Falciano:1986wk,Guanziroli:1987rp,Conway:1989fs,Anassontzis:1987hk}, 
with $A=p,d,W$, and $\bar{p} p \to \mu^+ \mu^- X$~\cite{Anassontzis:1987hk}, 
where the violation of the Lam-Tung sum rule strongly supports 
the conjecture to go beyond the collinear approximation~\cite{Conway:1989fs}. 

All this amount of puzzling measurements have triggered an intense theoretical
activity, particularly about the idea that intrinsic transverse momenta of
partons, together with transverse spin degrees of freedom, could be responsible
for the observed asymmetries. Transverse-Momentum Dependent (TMD) parton
distributions and fragmentation functions have been introduced and linked to
measurable asymmetries in the leading-twist cross sections of Semi-Inclusive DIS
(SIDIS), Drell-Yan process, semi-inclusive hadron-hadron collision, and $e^+ e^-$
annihilation~\cite{Ralston:1979ys,Collins:1981uk,Collins:1981uw,Collins:1984kg,Sivers:1990cc,Collins:1993kk,Kotzinian:1995dv,Anselmino:1995tv,Mulders:1996dh,Boer:1998nt}. 
As for parton distributions, the prototype of such TMD functions is the Sivers 
function~\cite{Sivers:1990cc}, which has a probabilistic interpretation: it 
describes how the distribution of unpolarized quarks is distorted by the 
transverse polarization of the parent hadron. Using the notations recommended in 
Ref.~\cite{Bacchetta:2004jz}, the Sivers function $f_{1T}^\perp$ can be
extracted by measuring the socalled Sivers effect in hadron-hadron collisions or 
SIDIS processes, i.e. an asymmetric distribution of the final-state products in 
the azimuthal angle defined by the mixed product ${\bm p}_\sT \times {\bm P}\cdot 
{\bm S}_\sT$, where ${\bm P}$ is the nucleon momentum and ${\bm p}_\sT, 
{\bm S}_\sT$ are the transverse components of the parton momentum and of the 
nucleon spin with respect to the direction of ${\bm P}$ in the infinite momentum 
frame. Time-reversal invariance would forbid such correlation if there were no 
initial/final-state interactions in the considered collision/SIDIS process, 
respectively. Therefore, $f_{1T}^\perp$ is conventionally named a "naive 
time-reversal-odd" distribution. The interactions must imply an interference 
between different helicity states of the target 
nucleon~\cite{Brodsky:2002cx,Ji:2002xn}; consequently, the correlation between 
${\bm p}_\sT$ and ${\bm S}_\sT$ is possible only for a nonvanishing orbital 
angular momentum of the partons. Then, extraction of $f_{1T}^\perp$ from data 
allows to study the orbital motion of hidden confined partons; better, it 
contains information on their spatial distribution~\cite{Burkardt:2003je}, and it 
offers a natural link between microscopic properties of confined elementary 
constituents and hadronic measurable quantities, such as the nucleon anomalous 
magnetic moment~\cite{Burkardt:2005km}.

In QCD, the necessary interactions can be naturally identified with the multiple
exchange of soft gluons contained in the gauge link operator, which grants the
color gauge-invariant definition of TMD 
distributions~\cite{Collins:2002kn,Ji:2002aa}. However, the whole picture relies 
on the proof of a suitable factorization theorem at small trasverse momenta for 
the process at hand. At present, QCD factorization proofs have been established 
for $e^+ e^-$ annihilations~\cite{Collins:1981uk}, for Drell-Yan 
processes~\cite{Collins:1984kg}, and, more recently, for SIDIS 
processes~\cite{Collins:2004nx,Ji:2004wu}, including also naive T-odd 
contributions. The related universality of TMD functions has been carefully 
discussed in 
Refs.~\cite{Collins:2002kn,Boer:2003cm,Collins:2004nx}. It turns 
out that the Sivers function displays the very interesting property of changing 
sign when going from the SIDIS to the Drell-Yan process, due to a peculiar 
feature of its gauge link operator under the time-reversal 
operation~\cite{Collins:2002kn}. This interesting prediction has stimulated 
intense experimental and phenomenological activities to link the Sivers effect 
recently measured at HERMES~\cite{Airapetian:2004tw,Diefenthaler:2005gx} with 
processes happening at RHIC, where data are being taken for polarized $pp$
collisions~\cite{Adler:2005in}. In particular, three different parametrizations 
of $f_{1T}^\perp$~\cite{Anselmino:2005ea,Vogelsang:2005cs,Collins:2005wb} have 
been extracted from the HERMES data (and found compatible also with the recent 
COMPASS data~\cite{Alexakhin:2005iw}), and have been used then to make 
predictions for SSA at RHIC (see also the more recent analysis of 
Ref.~\cite{Collins:2005rq}; for a comparison among the various approaches, see 
also Ref.~\cite{Anselmino:2005an}). 

In view of the foreseen upgrade of RHIC detector and luminosity (RHIC II), we will 
consider here the specific Drell-Yan process $pp^\uparrow \to \mu^+ \mu^- X$. 
The leading-twist polarized part of the cross section contains two terms that 
produce interesting SSA with azimuthal distinct behaviours~\cite{Boer:1999mm}. 
In a previous paper~\cite{Bianconi:2004wu}, we analyzed the term weighted by 
$\sin (\phi + \phi_S)$, with $\phi$ and $\phi_S$ the azimuthal orientations of 
the final lepton plane and of the proton polarization with respect to the 
reaction plane; it leads to the extraction of another interesting naive T-odd 
TMD distribution, the Boer-Mulders $h_1^\perp$, which is most likely responsible 
for the above mentioned violation of the Lam-Tung sum rule~\cite{Boer:1999mm}. 
In Ref.~\cite{Bianconi:2004wu}, we considered the Drell-Yan process
$\bar{p}p^\uparrow \to \mu^+ \mu^- X$ at the kinematics of interest for the High
Energy Storage Ring (HESR) project at GSI~\cite{Maggiora:2005cr,pax2} and we 
numerically simulated the SSA with a Monte Carlo in order to explore the minimal 
conditions required for an unambiguous extraction of $h_1^\perp$. Here, we 
follow the same approach to isolate the other term weighted by  $\sin (\phi - 
\phi_S)$, which contains the convolution of $f_{1T}^\perp$ with the standard 
unpolarized parton distribution $f_1$. The Monte Carlo will be applied to the 
$pp^\uparrow \to \mu^+ \mu^- X$ process at $\sqrt{s}=200$ GeV at the RHIC II luminosity
(at least $10^{32}$ cm$^{-2}$ sec$^{-1}$). The SSA will be numerically simulated using 
as input both the parametrization of Ref.~\cite{Anselmino:2005ea} and a new high-energy 
parametrization of $f_{1T}^\perp$ constrained by recent RHIC data on SSA for the
$pp^\uparrow \to \pi X$ process at the same $\sqrt{s}=200$ GeV~\cite{Adler:2005in}. The 
goal is to explore the sensitivity of the simulated asymmetry to different input 
parametrizations, as well as to directly verify, within the reached statistical 
accuracy, the predicted sign change of the Sivers function between SIDIS and 
Drell-Yan~\cite{Collins:2002kn}.

In Sec.~\ref{sec:mc}, the general formalism and details of the numerical 
simulation are briefly reviewed. In Sec.~\ref{sec:input}, we discuss the input 
parametrizations. In Sec.~\ref{sec:out}, results are presented. Finally, in 
Sec.~\ref{sec:end} some conclusions are drawn.

%%%%%%%%%%%%%%%%%%%%%%%%%% Fig. 1 %%%%%%%%%%%%%%%%%%%%%%%%%%%%%
\begin{figure}[h]
\centering
\includegraphics[width=7cm]{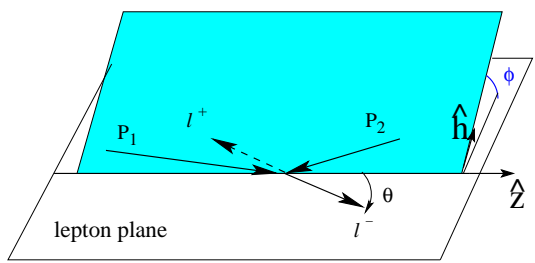}
\caption{The Collins-Soper frame.}
\label{fig:dyframe}
\end{figure}
%%%%%%%%%%%%%%%%%%%%%%%%%%%%%%%%%%%%%%%%%%%%%%%%%%%%%%%%%%%%%%%

%%%%%%%%%%%%%%%%%%%%%%%%%%%%%%%%%%%%%%%%%%%%%%%%%%%%%%%%%%%%
\section{Theoretical framework and numerical simulations}
\label{sec:mc}

In a Drell-Yan process, a lepton with momentum $k_1$ and an antilepton with
momentum $k_2$ (with $k_{1(2)}^2 \sim 0$) are produced from the collision of two
hadrons with momentum $P_1$, mass $M_1$, spin $S_1$, and $P_2, M_2, S_2$,
respectively (with $P_{1(2)}^2=M_{1(2)}^2, \; S_{1(2)}^2=-1, \; P_{1(2)}\cdot
S_{1(2)}=0$). The center-of-mass (cm) square energy available is $s=(P_1+P_2)^2$ 
and the invariant mass of the final lepton pair is given by the time-like 
momentum transfer $q^2 \equiv M^2 = (k_1 + k_2)^2$. In the kinematical regime 
where $M^2,s \rightarrow \infty$, while keeping the ratio $0\leq \tau = M^2/s 
\leq 1$ limited, the lepton pair can be assumed to be produced from the 
elementary annihilation of a parton and an antiparton with momenta $p_1$ and 
$p_2$, respectively. If $P_1^+$ and $P_2^-$ are the dominant light-cone 
components of hadron momenta in this regime, then the partons are approximately 
collinear with the parent hadrons and carry the light-cone momentum fractions 
$0\leq x_1 = p_1^+ / P_1^+ , \; x_2 = p_2^- / P_2^- \leq 1$, with $q^+ = p_1^+, 
\; q^- = p_2^-$ by momentum conservation~\cite{Boer:1999mm}. As already 
anticipated in Sec.~\ref{sec:intro}, a key issue is the extraction of TMD parton 
distributions; this requires the cross section to be kept differential in the 
transverse momentum of the final lepton pair, ${\bm q}_\sT$, which is bounded 
by the momentum conservation ${\bm q}_\sT = {\bm p}_{1\sT} + {\bm p}_{2\sT}$ to
each intrinsic transverse components ${\bm p}_{i\sT}$ of the parton momentum 
$p_i$ with respect to the direction defined by the corresponding hadron momentum 
${\bm P}_i$. If ${\bm q}_\sT \neq 0$ the annihilation direction is not known. 
Hence, it is convenient to select the socalled Collins-Soper 
frame~\cite{Collins:1977iv} (see fig.~\ref{fig:dyframe}), where
\bea
\hat{t} &= &\frac{q}{Q} \nn \\
\hat{z} &= &\frac{x_1 P_1}{Q} - \frac{x_2 P_2}{Q} \nn \\
\hat{h} &= &\frac{q_\sT}{|{\bm q}_\sT|} \; .
\label{eq:colsop-frame}
\eea
The final lepton pair is detected in the solid angle $(\theta, \phi )$, where
$\theta$ is defined in Fig.~\ref{fig:dyframe} and $\phi$ (and all other azimuthal 
angles) is measured in a plane perpendicular to $\hat{z}, \hat{t}$, but containing 
$\hat{h}$.

The full expression of the leading-twist differential cross section for the $H_1
H_2^\uparrow \to l^+ l^- X$ process can be written as~\cite{Boer:1999mm}
\bea
\frac{d\sigma}{d\Omega dx_1 dx_2 d{\bm q}_\sT} &= &
\frac{d\sigma^o}{d\Omega dx_1 dx_2 d{\bm q}_\sT} + 
\frac{d\Delta \sigma^\uparrow}{d\Omega dx_1 dx_2 d{\bm q}_\sT}  \nn \\
&= &\frac{\alpha^2}{3Q^2}\,\sum_f\,e_f^2\,\Bigg\{ A(y) \, 
{\cal F}\left[ f_1^f(H_1)\, f_1^f (H_2) \right] \nn \\
& &\mbox{\hspace{2cm}} + B(y) \, \cos 2\phi \, 
{\cal F}\left[ \left( 2 \hat{\bm h}\cdot {\bm p}_{1\sT} \, \hat{\bm h} \cdot 
{\bm p}_{2\sT} - {\bm p}_{1\sT} \cdot {\bm p}_{2\sT} \right) \, 
\frac{h_1^{\perp\,f}(H_1)\,h_1^{\perp\,f}(H_2)}{M_1\,M_2}\,\right] \Bigg\} \nn \\
& &+ \frac{\alpha^2}{3Q^2}\,|{\bm S}_{2\sT}|\,\sum_f\,e_f^2\,\Bigg\{ 
A(y) \, \sin (\phi - \phi_{S_2})\, {\cal F}\left[ \hat{\bm h}\cdot {\bm p}_{2\sT} 
\,\frac{f_1^f(H_1) \, f_{1_T}^{\perp\,f}(H_2)}{M_2}\right] \nn \\
& &\mbox{\hspace{3cm}} - B(y) \, \sin 
(\phi + \phi_{S_2})\, {\cal F}\left[ \hat{\bm h}\cdot {\bm p}_{1\sT} \,
\frac{h_1^{\perp\,f}(H_1) \, h_1^f(H_2)}{M_1}\right] \nn \\
& &\mbox{\hspace{3cm}}  - B(y) \, \sin (3\phi - \phi_{S_2})\, {\cal F}\left[ 
\left( 4 \hat{\bm h}\cdot {\bm p}_{1\sT} \, (\hat{\bm h} \cdot {\bm p}_{2\sT})^2 - 
2 \hat{\bm h} \cdot {\bm p}_{2\sT} \, {\bm p}_{1\sT} \cdot {\bm p}_{2\sT} - 
\hat{\bm h}\cdot {\bm p}_{1\sT} \, {\bm p}_{2\sT}^2 \right) \right. \nn \\
& &\mbox{\hspace{7cm}} \left. \times
\frac{h_1^{\perp\,f}(H_1) \, h_{1T}^{\perp\,f}(H_2)}{2 M_1\,M_2^2}\,\right]\, 
\Bigg\} \; ,
\label{eq:xsect}
\eea
where $\alpha$ is the fine structure constant, $d\Omega = \sin \theta d\theta
d\phi$, $e_f$ is the charge of the parton with flavor $f$, $\phi_{S_i}$ is
the azimuthal angle of the transverse spin of hadron $i$, and 
\begin{align}
A(y) = \left( \frac{1}{2} - y + y^2 \right) \, \stackrel{\mbox{cm}}{=}\, 
\frac{1}{4}\left( 1 + \cos^2 \theta \right) &\mbox{\hspace{2cm}} 
B(y) = y (1-y) \, \stackrel{\mbox{cm}}{=}\,\frac{1}{4}\, \sin^2 \theta \; . 
\label{eq:lepton}
\end{align}
The TMD functions $f_1^f(H), \, h_1^{\perp\,f}(H)$, describe the distributions of
unpolarized and transversely polarized partons in an unpolarized hadron $H$,
respectively, while $f_{1T}^{\perp\,f}(H), \, h_{1T}^{\perp\,f}(H)$, have a 
similar interpretation but for transversely polarized hadrons $H^\uparrow$. The 
transversity $h_1^f$ describes transversely polarized partons in transversely 
polarized hadrons. Each one of these distributions for a parton $f$ is convoluted 
with its antiparton partner $\bar{f}$ according to
\be
{\cal F} \left[ A^f(H_1) \, A^f(H_2) \right] \equiv \int d{\bm p}_{1\sT} d{\bm
p}_{2\sT}\, \delta \left( {\bm p}_{1\sT} + {\bm p}_{2\sT} - {\bm q}_\sT \right) \, 
\left[ A(x_1,{\bm p}_{1\sT}; \bar{f}/H_1)\, A(x_2,{\bm p}_{2\sT}; f/H_2^\uparrow ) 
+ (f\leftrightarrow \bar{f}) \right] \; .
\label{eq:convol}
\ee

In previous papers, we analyzed the SSA generated by the azimuthal dependences 
$\cos 2\phi$ and $\sin (\phi + \phi_{S_2})$ in 
Eq.~(\ref{eq:xsect})~\cite{Bianconi:2004wu}, as  well as the double-polarized 
Drell-Yan process~\cite{Bianconi:2005bd}. A combined measurement of these SSA 
allows to completely determine the unknown transversity $h_1$ and Boer-Mulders 
function $h_1^\perp$, which could be responsible for the well known violation of 
the Lam-Tung sum rule~\cite{Boer:1999mm} in unpolarized Drell-Yan 
data~\cite{Falciano:1986wk,Guanziroli:1987rp,Conway:1989fs} (see also 
Ref.~\cite{Boer:2005en}, and references therein, for a recent discussion on a
parallel with QCD vacuum effects). We set up a Monte Carlo simulation of Drell-Yan
processes involving unpolarized antiproton beams and transversely polarized 
protons for various kinematic scenarios at the HESR at GSI, namely for $30<s<200$ 
GeV$^2$ with an antiproton beam energy of 15 GeV and the socalled
asymmetric collider mode (proton beams of $~3$ GeV) or fixed target mode (proton
fixed targets). Special focus was put on the range $4<M<9$ GeV for the lepton
invariant mass, since it does not overlap with the charmonium and bottonium
resonances (where the elementary annihilation does not necessarily proceed through
a simple intermediate virtual photon) and higher-twist corrections should be
suppressed justifying the simple approach of Eq.~(\ref{eq:xsect}) based on the
parton model~\cite{Bianconi:2004wu}. Here, we will concentrate on the term 
weighted by $\sin (\phi - \phi_{S_2})$ in Eq.~(\ref{eq:xsect}) for $H_1=p$ and
$H_2^\uparrow = p^\uparrow$, and we will consider the related SSA at $\sqrt{s}=200$ GeV
reachable at RHIC~\cite{Bunce:2000uv}. Most 
of the technical details of the simulation are mutuated from our previous works; 
hence, we will heavily refer to Refs.~\cite{Bianconi:2004wu,Bianconi:2005bd} in 
the following.

The Monte Carlo events have been generated by the following cross 
section~\cite{Bianconi:2004wu}:
\be
\frac{d\sigma}{d\Omega dx_1 dx_2 d{\bm q}_\sT} = K \, \frac{1}{s}\, 
A({\bm q}_\sT, x_1, x_2, M)\,F(x_1, x_2)\, \sum_{i=1}^4\, c_i ({\bm q}_\sT, 
x_1,x_2) \, S_i(\theta, \phi, \phi_{_{S_2}}) \; .
\label{eq:mc-xsect}
\ee
It means that in Eq.~(\ref{eq:xsect}) we assume a factorized transverse-momentum
dependence in each parton distribution such as to break the convolution ${\cal
F}$, leading to the product $A\, F$. The function $A$ is parametrized 
as~\cite{Bianconi:2004wu}
\be
A(q_\sT,x_1,x_2,M) = \frac{5\,\displaystyle{\frac{a}{b}\,\left[ \frac{q_\sT}{b}
\right]^{a-1}}}{\left[ 1 + \left(\displaystyle{\frac{q_\sT}{b}}\right)^a 
\right]^6} \; ,
\label{eq:mcqT}
\ee
where $a(x_{_F},M),\,b(x_{_F},M),$ are parametric polynomials given in Appendix A 
of Ref.~\cite{Conway:1989fs} with $x_{_F} = x_1 - x_2$ and $q_\sT = 
|{\bm q}_\sT|$ (see also the more recent Ref.~\cite{Towell:2001nh}). It is 
normalized as
\be
\int dq_\sT \, A(q_\sT,x_1,x_2,M) = 1 \; .
\label{eq:mcqTnorm}
\ee
Actually, the Drell-Yan events studied in Ref.~\cite{Conway:1989fs} were produced 
for $\pi - p$ collisions; however, the same analysis, repeated for $\bar{p}-p$ 
and $p-p$ collisions~\cite{Anassontzis:1987hk}, gives a similar distribution for 
$q_\sT$ not very close to 0 and not much larger than 3 GeV/c. Here, we will 
adopt two different cuts on $q_\sT$ depending on the input parametrization for 
the Sivers function $f_{1T}^\perp$ (see next Sec.~\ref{sec:input}), namely 
$1<q_\sT<3$ GeV/$c$ and $0.1<q_\sT<2$ GeV/$c$. Anyway, the average transverse momentum 
turns out to be $\langle q_\sT \rangle > 1$ GeV/$c$, i.e. much bigger than parton 
intrinsic transverse momenta induced by confinement. 

The latter observation implies that sizeable QCD corrections affect the simple 
parton model picture of Eq.~(\ref{eq:xsect}). Their influence on the ${\bm q}_\sT$
distribution is effectively contained in the phenomenological parametrization of
Eq.~(\ref{eq:mcqT}). However, there are also other well known 
corrections~\cite{Altarelli:1979ub} coming from the resummation of leading 
logarithms at any order in the strong coupling constant $\alpha_s$, and from the 
inclusion of diagrams at first order in $\alpha_s$ involving the $\bar{q}q$ fusion
or the $q\g$ Compton mechanisms. The first group, usually named leading-log 
approximation (LLA), introduces a logarithmic dependence on the scale $M^2$ inside 
the various parameters entering the parton distributions~\cite{Buras:1977yj} 
contained in Eq.~(\ref{eq:mc-xsect}), such that it would determine their DGLAP 
evolution. However, the range of $M$ values here explored is close to the one of 
Refs.~\cite{Conway:1989fs,Anassontzis:1987hk}, where the parametrization of 
$A, F,$ and $c_i$ in Eq.~(\ref{eq:mc-xsect}) was deduced assuming $M$-independent 
parton distributions. Moreover, as it will be shown in the next Sec.~\ref{sec:out},
most of the events concentrate around the average $\langle x \rangle \sim 0.1$,
where the effects of evolution are almost vanishing. Therefore, similarly to 
Ref.~\cite{Bianconi:2004wu,Bianconi:2005bd} we take 
\be
F(x_1,x_2) = \frac{\alpha^2}{12 Q^2}\,\sum_f\,e_f^2\,
f_1^f(x_1; \bar{f}/H_1) \, f_1^f (x_2; f/H_2) + (\bar{f} \leftrightarrow f) \; , 
\label{eq:mcF}
\ee
which represents the azimuthally symmetric unpolarized part of 
Eq.~(\ref{eq:xsect}) that has been factorized out for convenience. As previously
stressed, the unpolarized distribution $f_1^f (x)$ for various flavors $f=u,d,s$, 
is parametrized as in Ref.~\cite{Anassontzis:1987hk}. 

The second group of QCD corrections is named next-to-leading-log approximation
(NLLA), and it is responsible for the well known $K$ factor in
Eq.~(\ref{eq:mc-xsect}). The $K$ factor is roughly independent on $x_{_F}$ and 
$M^2$ but it can grow with $\sqrt{\tau}$; it also depends on the chosen 
normalization of the parton distributions (for a detailed analysis in this context 
see Ref.~\cite{Anassontzis:1987hk}). It is a large correction, typically a 
multiplicative factor in the range $1.5 \div 2.5$. Here, we will conventionally 
assume the same value 2.5 adopted in our previous simulations.  But we stress 
that in an azimuthal asymmetry the corrections to the cross sections in the 
numerator and in the denominator should compensate each other. This is certainly 
true for each elementary contribution to the amplitude for SSA, but it is much 
less obvious for the ratio of full differential cross sections. Indeed, the 
smooth dependence of the SSA on NLLA corrections has been confirmed for fully 
polarized Drell-Yan processes at RHIC cm square energies~\cite{Martin:1998rz}. 

The whole solid angle $(\theta, \phi)$ of the final lepton pair in the 
Collins-Soper frame is randomly distributed in each variable. The explicit form 
for sorting the angular distribution in the Monte-Carlo 
is~\cite{Bianconi:2004wu,Bianconi:2005bd}
\bea
\sum_{i=1}^4\, c_i (q_\sT,x_1,x_2) \, S_i(\theta, \phi, \phi_{S_2}) &= 
&1 + \cos^2 \theta + \frac{\nu (x_1,x_2,q_\sT)}{2}\, \sin^2\theta \, \cos 2\phi 
\nn \\
& &+ |{\bm S}_{2\sT}|\, c_4 (q_\sT,x_1,x_2)\, S_4 (\theta, \phi, \phi_{S_2}) \; .
\label{eq:mcS}
\eea
If quarks were massless, the virtual photon would be only transversely polarized 
and the angular dependence would be described by the functions $c_1 = S_1 = 1$ 
and $c_2 = 1, \, S_2 = \cos^2 \theta$. Violations of such azimuthal symmetry 
induced by the function $c_3 \equiv \textstyle{\frac{\nu}{2}}$ are due to the 
longitudinal polarization of the virtual photon and to the fact that quarks have 
an intrinsic transverse momentum distribution, leading to the explicit violation
of the socalled Lam-Tung sum rule~\cite{Conway:1989fs}. QCD corrections influence 
$\nu$, which in principle depends also on $M^2$ (see App.~A of 
Ref.~\cite{Conway:1989fs}). Azimuthal $\cos 2\phi$ asymmetries induced by $\nu$ 
were simulated in Ref.~\cite{Bianconi:2004wu} using the simple parametrization of
Ref.~\cite{Boer:1999mm} and testing it against the previous measurement of
Ref.~\cite{Conway:1989fs}. 

The last term in Eq.~(\ref{eq:mcS}) corresponds to the polarized part of the 
cross section~(\ref{eq:xsect}). Since we want to single out just the Sivers
contribution, we assume that 
\be
S_4 (\theta, \phi, \phi_{S_2}) = (1+\cos^2 \theta) \, \sin (\phi - \phi_{S_2}) 
\; . 
\label{eq:mcS4}
\ee
Recalling that in Eq.~(\ref{eq:mc-xsect}) the azimuthally symmetric unpolarized 
part $A(q_\sT,x_1,x_2,M) \, F(x_1,x_2)$ of the cross section has been factorized 
out, the corresponding coefficient $c_4$ in Eq.~(\ref{eq:mcS}) in principle reads
\be
c_4 (q_\sT,x_1,x_2) = {\bm S}_{2\sT}\, 
\frac{\sum_f\,e_f^2\,{\cal F}\left[ \hat{\bm h}\cdot {\bm p}_{2\sT} \,
\displaystyle{\frac{f_1^f(x_1,{\bm p}_{1\sT})\, 
                    f_{1T}^{\perp\, f}(x_2,{\bm p}_{2\sT})}{M_2}} \right]}
{\sum_f\,e_f^2\,{\cal F}\left[ f_1^f(x_1,{\bm p}_{1\sT})\, 
 f_1^f(x_2,{\bm p}_{2\sT}) \right]} \; ,
\label{eq:mcc4}
\ee
where the complete dependence of the involved TMD parton distributions has been
made explicit. In the next Sec.~\ref{sec:input}, we will discuss two different
parametrizations for the $x$ and ${\bm p}_\sT$ dependence of these distributions
which allow to calculate the convolutions and determine $c_4$. 

Following Refs.~\cite{Bianconi:2004wu,Bianconi:2005bd}, the SSA corresponding to
the Sivers effect is constructed by dividing the event sample in two groups, one 
for positive values of $\sin (\phi - \phi_{S_2})$ ($U$) and another one for 
negative values ($D$), and taking the ratio $(U-D)/(U+D)$. Data are accumulated
only in the $x_2$ bin, i.e. they are summed over in $x_1, \theta,$ and in 
$q_\sT$ with the discussed cutoffs. Contrary to 
Refs.~\cite{Bianconi:2004wu,Bianconi:2005bd}, no cutoff has been applied to the
$\theta$ distribution because $S_4$ in Eq.~(\ref{eq:mcS4}) contains the term
$1+\cos^2 \theta$. Statistical errors for $(U-D)/(U+D)$ are obtained by 
making 10 independent repetitions of the simulation for each individual case, and 
then calculating for each $x_2$ bin the average asymmetry value and the variance. 
We checked that 10 repetitions are a reasonable threshold to have stable numbers, 
since the results do not change significantly when increasing the number of 
repetitions beyond 6. In a real experiment, the SSA would be extracted by taking 
the ratio between proper differences and sums of cross sections for the four 
possible combinations with the azimuthal angles $\pm \phi, \, \pm \phi_{S_2}$, in 
order to reduce systematic errors. In the Monte Carlo simulation, for each 
$\phi_{S_2}$ we can simply build the SSA in the $\phi$ angle. In an ideal 
experiment, the two situations would be equivalent. It is worth noting that while 
$\phi_{S_2}$ is fixed in the lab frame, in the Collins-Soper frame of 
Fig.~\ref{fig:dyframe} it is variable, since the $\hat{h}$ axis is directed along 
${\bm q}_\sT/q_\sT$; hence, a random distribution in $\phi_{S_2}$ must be 
initially extracted in the Monte Carlo.

%%%%%%%%%%%%%%%%%%%%%%%%%%%%%%%%%%%%%%%%%%%%%%%%%%%

\section{Parametrizations of the Sivers function}
\label{sec:input}

In our previous papers~\cite{Bianconi:2004wu,Bianconi:2005bd}, the strategy of the
numerical simulation was based on making guesses for the input $x$ and ${\bm
p}_\sT$ dependence of the parton distributions, and on trying to determine the
minimum number of events required to discriminate various SSA produced by very
different input guesses. In fact, this would be equivalent to state that in this
case some analytic information on the structure of these TMD parton distributions
could be extracted from the SSA measurement. 

As for the Sivers effect, the situation is different because recently the HERMES 
collaboration has released new SSA data for the SIDIS process on transversely
polarized protons~\cite{Diefenthaler:2005gx}, which substantially increase the
precision of the previous data set~\cite{Airapetian:2004tw}. As a consequence, 
three different parametrizations of 
$f_{1T}^\perp$~\cite{Anselmino:2005ea,Vogelsang:2005cs,Collins:2005wb} have 
been extracted from this data set and found compatible also with the recent 
COMPASS data~\cite{Alexakhin:2005iw}. Moreover, a recent preprint appeared which
usefully illustrates the differences among the various 
approaches~\cite{Anselmino:2005an}. At the same time, new data have been
collected at RHIC~\cite{Adler:2005in,Videbaek:2005fm} on SSA in the $pp^\uparrow
\to \pi X$ process, that confirm the observation of large asymmetries at forward
rapidity of the pion also at the high-energy collider regime ($\sqrt{s}=200$ GeV). 
Despite this class of hadron-hadron collisions is power suppressed and factorization 
was established in terms of higher-twist correlation operators~\cite{Qiu:1991pp}, 
still the SSA receives a contribution from the leading-twist convolution $f_1 \otimes
f_{1T}^\perp \otimes D_1$, where $D_1$ describes the fragmentation of an
unpolarized quark into the detected $\pi$. Therefore, analogously to the analysis of
Ref.~\cite{D'Alesio:2004upky} at lower energy~\cite{Adams:1991cs}, we believe that the 
measured SSA in $pp^\uparrow \to \pi X$ processes at $\sqrt{s}=200$ GeV should 
indirectly constrain the parametrization of $f_{1T}^\perp$ and, consequently, the 
"strength" of the Sivers effect when these data are ideally interpreted as completely 
driven by the above convolution. 

In our Monte Carlo simulation, we consider two different parametrizations for
$f_{1T}^\perp$: the one elaborated in Ref.~\cite{Anselmino:2005ea} and based on
the new HERMES~\cite{Diefenthaler:2005gx} and COMPASS~\cite{Alexakhin:2005iw}
data, where the ${\bm p}_\sT$ dependence is driven by the $\langle {\bm p}_\sT^2
\rangle$ extracted in a model dependent way from the azimuthal asymmetry of the 
unpolarized SIDIS cross section (Cahn effect); a new high-energy parametrization 
inspired to the one of Ref.~\cite{Vogelsang:2005cs} but with a specific 
${\bm p}_\sT$ dependence constrained by the $pp^\uparrow \to \pi X$ data at
$\sqrt{s}=200$ GeV.

%%%%%%%%%%%%%%%%%%%%%%%%%%%%%%%%%%%%%%%%%%%%%%%%%%%

\subsubsection{The parametrization of Ref.~\cite{Anselmino:2005ea}}
\label{sec:anselmino}

Keeping in mind the commonly adopted conventions~\cite{Bacchetta:2004jz}, the
expression used in Ref.~\cite{Anselmino:2005ea} is 
\bea
f_{1T}^{\perp\, f}(x,{\bm p}_\sT) &=&- \frac{M_2}{2 p_\sT}\, \Delta^N
f_{f/p^\uparrow} (x, {\bm p}_\sT) \nn \\
&= &-2\, N_f\,\frac{(a_f+b_f)^{a_f+b_f}}{a_f^{a_f}\,b_f^{b_f}}\,
x^{a_f}\,(1-x)^{b_f}\,\frac{M_2 M_0}{{\bm p}_\sT^2+M_0^2}\,
f_1^f(x,{\bm p}_\sT) \nn \\
&= &-2\, N_f\,\frac{1}{\pi \, \langle p_\sT^2 \rangle}\,
\frac{(a_f+b_f)^{a_f+b_f}}{a_f^{a_f}\, b_f^{b_f}} \, x^{a_f}\, (1-x)^{b_f}\, 
\frac{M_2 M_0}{{\bm p}_\sT^2+M_0^2}\, e^{-p_\sT^2/\langle p_\sT^2 \rangle}\, 
f_1^f(x) \; ,
\label{eq:pTanselm}
\eea
where $M_2$ is the mass of the polarized proton, $p_\sT \equiv |{\bm p}_\sT|$, 
and $\langle p_\sT^2 \rangle = 0.25$ (GeV/$c$)$^2$ is deduced by assuming a 
Gaussian form for the ${\bm p}_\sT$ dependence of $f_1$ in order to reproduce the 
azimuthal angular dependence of the SIDIS unpolarized cross section (Cahn 
effect). The parameters $N_f,a_f,b_f,$ with $f=u,d$ are extracted by fitting the 
recent HERMES~\cite{Diefenthaler:2005gx} and COMPASS~\cite{Alexakhin:2005iw} data 
with a final $\chi^2$ per degree of freedom of 1.06 (the negligible contribution 
of antiquarks in the minimization procedure has been traded off for a better 
precision). They are listed in Tab.~\ref{tab:pTanselm}.

%%%%%%%%%%%%%%%%%%%%%%%%%%%%%%%%%%%%%%%%%%%%%%%%%%%% 
\begin{table}[h]
\caption{\label{tab:pTanselm} Parameters for the Sivers distribution from
Ref.~\protect{\cite{Anselmino:2005ea}}}
\begin{ruledtabular}
\begin{tabular}{cccc}
quark up & {} & quark down & {} \\
\hline
$N_u$ & $0.32 \pm 0.11$ & $N_d$ & $-1.0 \pm 0.12$  \\
$a_u$ & $0.29 \pm 0.35$ & $a_d$ & $1.16 \pm 0.47$ \\
$b_u$ & $0.53 \pm 3.58$ & $b_d$ & $3.77 \pm 2.59$ \\
\hline
$M_0^2$ & $0.32 \pm 0.25$ (GeV/$c$)$^2$ & &  \\
\end{tabular}
\end{ruledtabular}
\end{table}
%%%%%%%%%%%%%%%%%%%%%%%%%%%%%%%%%%%%%%%%%%%%%%%%%%%%

Following the prediction about a sign change of $f_{1T}^\perp$ when going from a
SIDIS process (as for the HERMES analysis) to a Drell-Yan process (as it is
considered here), we insert the opposite of Eq.~(\ref{eq:pTanselm}) inside 
Eq.~(\ref{eq:mcc4}), including the Gaussian parametrization for 
$f_1(x,{\bm p}_\sT)$. The integrals upon the transverse momenta can be evaluated 
following the steps in Sec.VI of Ref.~\cite{Boer:1999mm}. The net result is
\be
c_4 \approx |{\bm S}_{2\sT}|\,\frac{4 M_0\,q_\sT}{q_\sT^2+4 M_0^2}\, 
\frac{\sum_f\,e_f^2\,N_f\, 
      \displaystyle{\frac{(a_f+b_f)^{a_f+b_f}}{a_f^{a_f}\,b_f^{b_f}} }\, 
      f_1(x_1;\bar{f}/p)\, x_2^{a_f}\,(1-x_2)^{b_f}\, f_1(x_2;f/p^\uparrow)}
     {\sum_f\,e_f^2\, f_1(x_1;\bar{f}/p)\, f_1^f(x_2;f/p)+ (1\leftrightarrow 2)} 
\; ,
\label{eq:c4-anselm}
\ee
We further simplify this expression by replacing the flavor-dependent product of
parton distributions with an average product $\langle f_1(x_1) \rangle \, \langle
f_1(x_2) \rangle$ both in the numerator and in the denominator, in order to reduce
the statistical noise related to the parametrization of $f_1(x)$. The final
expression of $c_4$ for $pp^\uparrow$ collisions, that numerically simulates the 
Sivers effect in our Monte Carlo, becomes~\footnote{In Eq.~(\ref{eq:c4mc-anselm}), the
factor in front of the flavor-dependent term should read $1/18$ because of the symmetry
operation in the denominator of Eq.~(\ref{eq:c4-anselm}). However, as it is shown in the
next Sec.~\ref{sec:scatter} the SSA is not suppressed only in the $(x_1<0,\,x_2>0)$
region of the phase space, which corresponds to take the dominant part of just the 
first term in the denominator of Eq.~(\ref{eq:c4-anselm}).}
\be
c_4 \approx |{\bm S}_{2\sT}|\,\frac{4 M_0\,q_\sT}{q_\sT^2+4 M_0^2}\, 
\frac{1}{9}\, \left[ 8\, N_u\, \frac{(a_u+b_u)^{a_u+b_u}}{a_u^{a_u}\,b_u^{b_u}} 
\, x_2^{a_u}\,(1-x_2)^{b_u}\, + \, N_d\, 
\frac{(a_d+b_d)^{a_d+b_d}}{a_d^{a_d}\,b_d^{b_d}} \, x_2^{a_d}\,(1-x_2)^{b_d} 
\right] \; .
\label{eq:c4mc-anselm}
\ee

As it is evident from previous formulae, the parametrization~(\ref{eq:pTanselm}) is
put in a very convenient form that easily fits the asymmetry term $c_4$ in our
Monte Carlo. However, the sometimes poor resolution in determining the parameters
forced us to select only the central values in Tab.~\ref{tab:pTanselm} in order to
produce meaningful numerical simulations. The sensitivity of the parameters to 
the HERMES results for the Sivers effect reflects in a more important relative 
weight of the $d$ quark over the $u$ one in the valence $x$ range, with opposite 
signs for the corresponding normalization $N_f, \, f=u,d$. As it will be shown in 
the next Sec.~\ref{sec:out}, this has two main consequences on the simulation: 
small SSA are obtained for the $pp^\uparrow \to \mu^+ \mu^- X$ process in the 
valence $x$ range, where the $\bar{d} d$ annihilation occurs less frequently than 
the $\bar{u} u$ one; a significant minimum number of events is necessary in the 
sample to reduce the statistical error bars and make the asymmetry not compatible 
with zero. From Tab.~\ref{tab:pTanselm}, the $a_u$ parameter is much smaller than 
1. This means that in Eq.~(\ref{eq:c4mc-anselm}) the $u$-quark term dominates at 
small $x_2$, leading to a persistence of the Sivers effect even below the valence 
$x$ range. This feature is potentially very relevant at RHIC kinematics, where 
$\langle x \rangle \sim 0.01$. Therefore, for this parametrization we have 
produced also a specific simulation a small $x_2$ with a finer binning 
$\Delta x_2$, as it will be shown in the next Sec.~\ref{sec:out}. The
flavor-independent Lorentzian shape in the ${\bm p}_\sT$ dependence of
Eq.~(\ref{eq:pTanselm}) produces a maximum asymmetry for $q_\sT \sim 1$ GeV/$c$
and a rapid decrease for larger values. Consequently, transverse momenta are
selected in the range $0.1< q_\sT < 2$ GeV/$c$, because for larger cutoffs the
asymmetry is diluted. 

%%%%%%%%%%%%%%%%%%%%%%%%%%%%%%%%%%%%%%%%%%%%%%%%%%%

\subsubsection{A new high-energy parametrization}
\label{sec:noi}

As already mentioned above, this new parametrization is inspired to the one of
Ref.~\cite{Vogelsang:2005cs}. There, it was assumed that the transverse momentum
of the detected pion in the SIDIS process was entirely due to the
transverse-momentum dependence in the Sivers function. No transverse momenta are 
contributed by other terms in the factorization formula. In this perspective, this
approach can be considered as a limiting case of the approach of 
Ref.~\cite{Collins:2005wb}, based on Gaussian ans\"atze for the ${\bm p}_\sT$
dependence both in the distribution and fragmentation functions (see 
Ref.~\cite{Anselmino:2005an} for a more detailed discussion). In 
Ref.~\cite{Vogelsang:2005cs}, no further assumption was made but the ${\bm p}_\sT$
distribution was integrated out. As a result, the SSA for the Sivers effect in
SIDIS was expressed in terms of "$\half$-moments" of the Sivers function, which
were parametrized in terms of the $u$-quark distribution $f_1^u(x)$ and
flavor-dependent normalizations $S_u, S_d,$ to be determined by a fit to the new
HERMES data~\cite{Diefenthaler:2005gx}. Also in this case the normalizations turn
out to have opposite sign, and the $\chi^2$ per degree of freedom is 1.2. The
recent COMPASS data~\cite{Alexakhin:2005iw} were not included in the fit, but a
direct comparison show a qualitative good agreement. 

Here, we retain the $x$ dependence of this approach, but we introduce a different
flavor-dependent normalization and an explicit ${\bm p}_\sT$ dependence that are
bound to the shape of the recent RHIC data on $pp^\uparrow \to \pi X$ at
$\sqrt{s}=200$ GeV~\cite{Adler:2005in}. The expression adopted is
\bea
f_{1T}^{\perp\, f}(x,{\bm p}_\sT) &=&N_f\,x\,(1-x)\,
\frac{M_2 p_0^2 p_\sT}{(p_\sT^2+\frac{p_0^2}{4})^2}\,f_1^f(x,{\bm p}_\sT) \nn \\
&= & N_f\,x\,(1-x)\,\frac{M_2 p_0^2 p_\sT}{(p_\sT^2+\frac{p_0^2}{4})^2}\,
\frac{1}{\pi \, \langle p_\sT^2 \rangle}\, e^{-p_\sT^2/\langle p_\sT^2 \rangle}\, 
f_1^f(x) \; ,
\label{eq:pTnoi}
\eea
where $p_0 = 2$ GeV/$c$. Following the same arguments of previous section, we get
\be
c_4 \approx |{\bm S}_{2\sT}|\,x_2 \, (1-x_2)\, 
\left( \frac{2\, p_0\, q_\sT}{q_\sT^2+p_0^2} \right)^2 \, 
\frac{\sum_f\,e_f^2\,N_f\, f_1(x_1;\bar{f}/p)\, f_1(x_2;f/p^\uparrow)}
     {\sum_f\,e_f^2\, f_1(x_1;\bar{f}/p)\, f_1^f(x_2;f/p)+ (1\leftrightarrow 2)} 
\; .
\label{eq:c4-noi}
\ee
Again, we can further simplify the expression introducing the flavor average 
product $\langle f_1(x_1) \rangle \, \langle f_1(x_2) \rangle$, which leads
to~\footnote{An argument similar to the one in the previous footnote about
Eq.~(\ref{eq:c4mc-anselm}), applies also here to Eq.~(\ref{eq:c4mc-noi}).}
\be
c_4 \approx |{\bm S}_{2\sT}|\, x_2 \, (1-x_2)\, 
\left( \frac{2\, p_0 \, q_\sT}{q_\sT^2+p_0^2} \right)^2 \, \frac{8\, N_u + N_d}{9} 
\; .
\label{eq:c4mc-noi}
\ee

The $q_\sT$ shape is different from Eq.~(\ref{eq:c4mc-anselm}) and the peak 
position is shifted at much larger values. This is in agreement with a similar 
analysis of the azimuthal asymmetry of the unpolarized Drell-Yan data (the 
violation of the Lam-Tung sum rule~\cite{Boer:1999mm}). But, more specifically, 
it is induced by the observed $x_{_F}-q_\sT$ correlation in the RHIC data for 
$pp^\uparrow \to \pi X$, when it is assumed that the SSA is entirely due to the
Sivers mechanism; this suggests that the maximum asymmetry is reached in 
the upper valence region such that $x_{_F} \approx x_2 \sim \langle q_\sT \rangle /
5$~\cite{Adler:2005in}. We have conveniently modified the cutoffs such that for
this parametrization the sampled distribution is $1< q_\sT < 3$ GeV/$c$. In this
case, the peak asymmetry is reached for $x_2 \sim 0.5$ (see next
Sec.~\ref{sec:out}). Contrary to the other Lorentzian and Gaussian distributions
adopted so far, the ${\bm p}_\sT$ distribution of Eq.~(\ref{eq:pTnoi}) cannot be
normalized in the usual way because of lack of convergence at large $q_\sT$.
Rather, it is built as to assume the value 1 on its peak position $q_\sT = p_0$.
Finally, the flavor dependence of the normalization is kept as simple as in 
Ref.~\cite{Collins:2005wb}, namely $N_u = - N_d = 0.7$. The sign, positive for $u$
quark and negative for the $d$, already takes into account the predicted sign
change of $f_{1T}^\perp$ from Drell-Yan to SIDIS, where the opposite flavor
dependence of the sign was 
obtained~\cite{Anselmino:2005ea,Vogelsang:2005cs,Collins:2005wb}.

%%%%%%%%%%%%%%%%%%%%%%%%%%%%%%%%%%%%%%%%%%%%%%%%%%%
\section{Results of the Monte Carlo simulations}
\label{sec:out}

In this Section, we present results for Monte Carlo simulations of the Sivers
effect in the Drell-Yan process $pp^\uparrow \to \mu^+ \mu^- X$ using input from
the previous Sec.~\ref{sec:anselmino} and \ref{sec:noi}. The goal is to explore 
the sensitivity of the simulated asymmetry to different input parametrizations 
(i.e. to the input theoretical uncertainty), as well as to directly verify, 
within the reached statistical accuracy, the predicted sign change of the Sivers 
function between SIDIS and Drell-Yan~\cite{Collins:2002kn}. The collision is
considered at the cm energy $\sqrt{s} = 200$ GeV, at which RHIC is presently
taking data. We use a conservative dilution factor 0.5 for the proton
polarization, even if a foreseen upgrade of the superconducting siberian snake 
will allow RHIC to run in the future with stable 70\% transverse 
polarization~\cite{Aidala:2005my}. We select two different ranges for the
lepton invariant mass: $4<M<9$ GeV and $12<M<40$ GeV. In this way, we avoid
overlaps with the resonance regions of the $\bar{c}c$ and $\bar{b}b$ quarkonium
systems. At the same time, the theoretical analysis based on the leading-twist
cross section~(\ref{eq:xsect}) should be well established, since higher-twist 
effects can be classified according to powers of $M_p/M$, where $M_p$ is the 
proton mass. Moreover, at $\sqrt{s}=200$ GeV also the QCD corrections beyond tree
level should be suppressed, which justifies the approximations described in
Sec.~\ref{sec:mc}. In the Monte Carlo, the events are sorted according to the
cross section~(\ref{eq:mc-xsect}), supplemented by
Eqs.~(\ref{eq:mcqT})-(\ref{eq:mcF}), while the asymmetry related to the Sivers
effect is simulated by Eqs.~(\ref{eq:mcS}), (\ref{eq:mcS4}), 
(\ref{eq:c4mc-anselm}) and (\ref{eq:c4mc-noi}). In particular, the events are
divided in two groups, one for positive values ($U$) of $\sin (\phi - 
\phi_{S_2})$ in Eq.~(\ref{eq:mcS4}), and another one for negative values ($D$), 
and taking the ratio $(U-D)/(U+D)$. Data are accumulated only in the $x_2$ bins of
the polarized proton, i.e. they are summed over in the $x_1$ bins for the
unpolarized proton, in the transverse momentum $q_\sT$ of the muon pair and in
their zenithal orientation $\theta$. Proper cuts are applied to the $q_\sT$
distribution according to the different input parametrization of the Sivers
function: for the case of Sec.~\ref{sec:anselmino}, $0.1<q_\sT <2$ GeV/$c$; for
the case of Sec.~\ref{sec:noi}, $1<q_\sT < 3$ GeV/$c$. In this way, the ratio
between the absolute sizes of the asymmetry and the statistical errors is
optimized for each choice. The resulting $\langle q_\sT \rangle$ is $\sim 1.8$ 
GeV/$c$, in fair agreement with the one experimentally explored at 
RHIC~\cite{Adler:2005in}. Contrary to 
Refs.~\cite{Bianconi:2004wu,Bianconi:2005bd}, there is no need to introduce cuts 
in the $\theta$ distribution because of the $(1+\cos^2 \theta)$ term in
Eq.~(\ref{eq:mcS4}). We have considered two initial different samples of 
$20\,000$ and $100\,000$ events. Statistical errors for $(U-D)/(U+D)$ are obtained by 
making 10 independent repetitions of the simulation for each individual case, and 
then calculating for each $x_2$ bin the average asymmetry value and the variance. 
We checked that 10 repetitions are a reasonable threshold to have stable numbers, 
since the results do not change significantly when increasing the number of 
repetitions beyond 6.

%%%%%%%%%%%%%%%%%%%%%%%%%% Fig. 2 %%%%%%%%%%%%%%%%%%%%%%%%%%%%%
\begin{figure}[h]
\centering
\includegraphics[width=7cm]{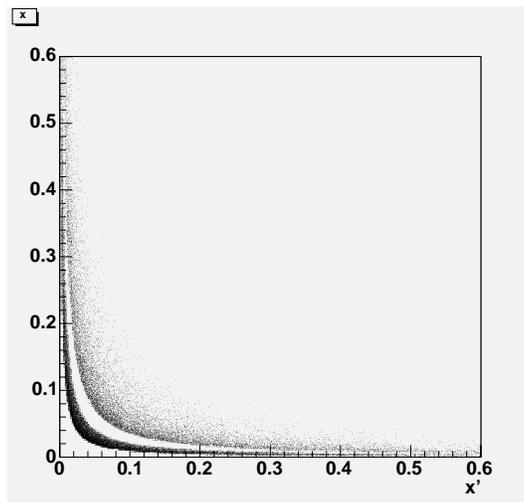}
%\vspace{7cm}
\caption{The scatter plot for $120\,000$ events of Drell-Yan muon pairs produced by
proton-proton collisions at $\sqrt{s}=200$ GeV. The two bands in which data are
grouped, correspond to two different ranges in the muon invariant mass: $4<M<9$ GeV
for the lower band, $12<M<40$ GeV for the upper one.}
\label{fig:scatter}
\end{figure}
%%%%%%%%%%%%%%%%%%%%%%%%%%%%%%%%%%%%%%%%%%%%%%%%%%%%%%%%%%%%%%%

%%%%%%%%%%%%%%%%%%%%%%%%%%%%%%%%%%%%%%%%%%%%%%%%%%

\subsection{Properties of the phase space}
\label{sec:scatter}

In Fig.~\ref{fig:scatter}, the scatter plot (in the fractional momenta $x_1\equiv
x', x_2\equiv x,$ of the annihilating partons) for $120\,000$ events of Drell-Yan 
muon pairs produced by proton-proton collisions at $\sqrt{s}=200$ GeV, is shown. 
The data are divided in 
two different bands, corresponding to two different ranges in the muon pair 
invariant mass: $4<M<9$ GeV for the lower band and $12<M<40$ GeV for the upper 
one. In fact, hyperboles $x_1 x_2 = const.$ are selected by fixed values of 
$\tau = x_1 x_2 = M^2/s$. Since the elementary annihilation is assumed to proceed 
through a virtual photon, the cross section contains a term $1/M^2 \sim 1/\tau$ 
which populates the phase space at low values, while the upper right corner of 
Fig.~\ref{fig:scatter} for $\tau \to 1$ is basically empty (also because the 
parton distributions vanish for $x_{1/2}\to 1$). Therefore, within each band 
events tend to accumulate to the lowest possible $M$ in the considered range, 
which means that they try to align along the hyperbole with lowest possible values 
of $x_1$ and $x_2$. Moreover, for the same reason the lower band is much more 
dense than the other one: 95\% of the events correspond to the $4<M<9$ GeV range. 
This is why we consider also this case, because the much higher statistics can be 
traded for the questionable neglect of higher-twist contributions with respect to 
the higher $12<M<40$ GeV range. 

The scatter plot of Fig.~\ref{fig:scatter} contains two main differences with
respect to our previous analysis of Ref.~\cite{Bianconi:2004wu} for the GSI setup, 
where the $\bar{p} p^\uparrow \to \mu^+ \mu^- X$ process was considered at 
$\sqrt{s}=14$ GeV. First of all, here the cm energy is higher by one order of 
magnitude, which means that $x_{1/2}$ values lower by one order of magnitude are 
explored, typically $\langle x_{1/2} \rangle \sim 0.01$. Secondly, this is 
emphasized by the fact that in $pp$ collisions at least one of the two 
annihilating partons comes from the proton sea distribution, which is peaked at 
very low $x$ values. The importance of parton momenta below the valence region is 
potentially relevant to the theoretical models. We already mentioned in 
Sec.~\ref{sec:anselmino} that the parametrization of Eq.~(\ref{eq:pTanselm}) leads 
to a persistence of the Sivers effect in this range, contrary to the other one of 
Eq.~(\ref{eq:pTnoi}). Anyway, both choices lead to a vanishing effect for $x_2 \to
0$. The dominance of the low $x_2$ portion of phase space is evident in the
histograms for the event distributions displayed in the next Sec.~\ref{sec:out},  
where the first bin $[0,0.1]$ contains more than 50\% of events on average. We can
also estimate the expected position of the peak density, assuming that the 
bidimensional event distribution $N(x_1,x_2)$ is dominated by the $1/\tau$ factor 
associated to the elementary $\bar{q} q$ fusion into a virtual photon. In this 
case, $N(x_1,x_2)=1/(x_1 x_2)$ for $\tau > M^2_{min}/s$, and 0 otherwise. 
Consequently, the $x_1$-integrated distribution has the form 
$\log (x_2 s/M^2_{min})/x_2$ and reaches its peak value for 
$\log (x_2 s/M^2_{min}) = 1$, i.e. for $x_2 = e M^2_{min}/s \approx 
3 M^2_{min} /s \sim 0.01$ for $\sqrt{s}=200$ GeV and $M_{min}=12$ 
GeV, in agreement with the average values produced by the Monte Carlo. Elsewhere, 
the integrated distribution behaves like $1/x_2$, as it can visually be checked in 
the histograms of next Sec.~\ref{sec:out}. 

The $1/\tau$ mechanism in the random generation can artificially suppress the
asymmetry irrespective of the size of the Sivers function itself. In fact, let us
consider the $12<M<40$ GeV range, which has a lower event rate; we distinguish four
different slices of phase space:
\begin{itemize}
\item{$x_1, x_2 > 0.1$ :} this part covers 99\% of the phase space, but it contains
only 0.5\% of the total number of events; it corresponds to higher $M$ values
($M>20$ GeV) and it is suppressed by the $1/M^2$ mechanism; moreover, in 
$F(x_1,x_2)$ of Eq.~(\ref{eq:mcF}) the annihilating antiquark with flavor 
$\bar{f}$ is picked up from the sea distribution of one of the two protons at 
large $x$;

\item{$x_1, x_2 < 0.1$ :} this part covers $< 1$\% of the phase space, but it 
contains 20\% of events; in fact, it corresponds to lower $M$ values (emphasized by
the $1/M^2$ mechanism) and $F(x_1,x_2)$ is dominated by the sea distributions in 
both protons, that are enhanced at small $x$; however, the SSA is suppressed 
because $f_{1T}^\perp / f_1 \to 0$ for $x_2 \to 0$;

\item{$x_1 > 0.1, x_2 < 0.1$ :} this part covers again $< 1$\% of the phase space,
but it contains 40\% of the events; it is less favoured by the $1/M^2$ mechanism
with respect to the previous case, but $F(x_1,x_2)$ contains the term 
$f_1(x_1;f/p)\, f_1(x_2;\bar{f}/p)$, that is dominant in this slice of phase 
space; however, for the very same reason the SSA is suppressed because it is 
approximately driven by $f_{1T}^\perp(x_2) / f_1(x_2;\bar{f}/p) \to 0$ for 
$x_2 \to 0$;

\item{$x_1 < 0.1, x_2 > 0.1$ :} all previous arguments apply also here upon the
$x_1 \leftrightarrow x_2$ exchange but for the SSA, which is driven by 
$f_{1T}^\perp(x_2) / f_1(x_2;f/p)$ and, therefore, it is not suppressed for $x_2 >
0.1$. 
\end{itemize}
In summary, irrespective of the size of the Sivers function, the magnitude of the
corresponding SSA is suppressed in all parts of phase space but in the region $x_1
< 0.1, x_2 > 0.1$, dominated by the sea partons of the unpolarized proton and by
the valence partons of the polarized one. Elsewhere, the $1/M^2$ mechanism induces
a dominance of the sea partons, that acts as an effective dilution factor leading
to a waste of $\sim 50$\% of the total number of events.

%%%%%%%%%%%%%%%%%%%%%%%%%%%%%%%%%%%%%%%%%%%%%%%%%%

\subsection{Total cross section and event rates}
\label{sec:rates}

Using Eq.~(\ref{eq:mcF}) with the parametrization for the parton distributions from
Ref.~\cite{Conway:1989fs}, from our Monte Carlo we deduce a total cross section
$\sigma_{pp}=0.1$ nb for the $pp\to \mu^+ \mu^- X$ process at $\sqrt{s}=200$ GeV
and with invariant masses in the range $12<M<40$ GeV, while we get
$\sigma_{pp}=1.2$ nb for the lower $4<M<9$ GeV range. The results are quite
sensitive to the parametrization of the parton distributions. We have recalculated
the total cross sections with the more recent NNLO analysis of 
Ref.~\cite{Martin:2002dr} and we get 0.4 nb and 7 nb, respectively. When changing
parametrizations and, consequently, normalizations, we had to readjust the $K$
factor accordingly; in order to reproduce the measured cross section at $\sqrt{s}
\approx 16$ GeV~\cite{Conway:1989fs}, we had to reduce it by 50\%. It means that
now $K \approx 1$ and the QCD corrections are mostly contained in the NNLO
parametrization of the parton distributions. The
sensitivity (and the related uncertainty) of this analysis to the input are 
sizeable, but do not alter the order of magnitude of the result. Since our goal is 
to estimate event rates by multiplying $\sigma_{pp}$ with a given luminosity, we 
are confident that the results are realistic and reliable. For RHIC, a luminosity
of $10^{32}$ cm$^{-2}$ sec$^{-1}$ or higher is foreseen~\cite{Bunce:2000uv}. This 
means, for example, that at least $250\,000$ Drell-Yan events/month (and up to 7 times 
more) could be collected with this luminosity and muon pair invariant masses in 
the $4<M<9$ GeV range. A list of the combinations here explored is given in 
Tab.~\ref{tab:rates} (for a more comprehensive analysis see 
Ref.~\cite{Bianconi:2005bv}).

%%%%%%%%%%%%%%%%%%%%%%%%%%%%%%%%%%%%%%%%%%%%%%%%%%%%%%

\begin{table}
\caption{\label{tab:rates} Total cross sections for Drell-Yan $pp$ collisions at 
$\sqrt{s}=200$ GeV and for various invariant masses of the muon pair, at the
given luminosity $10^{32}$ cm$^{-2}$ sec$^{-1}$.}
\begin{ruledtabular}
\begin{tabular}{ccccc}
$f_1(x)$ from & $M$ (GeV) & $\sigma_{pp}$ (nb) & rates (events/month) \\
\hline
Ref.~\cite{Conway:1989fs} & $4 \div 9$ & 1.2 & $2.5 \times 10^5$ \\
Ref.~\cite{Conway:1989fs} & $12 \div 40$ & 0.1 & $2.5 \times 10^4$ \\
Ref.~\cite{Martin:2002dr} & $4 \div 9$ & 7 & $1.5 \times 10^6$ \\
Ref.~\cite{Martin:2002dr} & $12 \div 40$ & 0.4 & $10^5$ \\
\end{tabular}
\end{ruledtabular}
\end{table}

%%%%%%%%%%%%%%%%%%%%%%%%%%%%%%%%%%%%%%%%%%%%%%%%%%%%%%

It is also interesting to compare with the antiproton-proton Drell-Yan collision.
In general, we expect that the lower the $\langle x \rangle$, the more the cross
sections are dominated by sea parton distributions, the closer the ratio
$\sigma_{\bar{p}p} / \sigma_{pp}$ approaches 1. By updating our previous
results~\cite{Bianconi:2004wu} at the present energy and viceversa, we get 
\begin{align}
\frac{\sigma_{\bar{p}p}}{\sigma_{pp}}(\sqrt{s}=200; 4<M<9) = 2 &\mbox{\hspace{1cm}}
\frac{\sigma_{\bar{p}p}}{\sigma_{pp}}(\sqrt{s}=200; 12<M<40) = 4 & 
\frac{\sigma_{\bar{p}p}}{\sigma_{pp}}(\sqrt{s}=14; 4<M<9) = 40 \; .
\label{eq:totratio}
\end{align}
There are two ways to lower the range of $x$ or, equivalently, $\tau$: decreasing 
the invariant mass $M$, or increasing the cm energy $\sqrt{s}$. Indeed, the 
results show that for this trend the ratio approaches 1. When increasing 
$\sqrt{s}$ at a given $M$ range, for example, the depletion of the ratio implies 
also that $\sigma_{pp}$ increases. This curious result of an increasing cross 
section with energy can be explained by recalling that a shift to smaller $x_1,
x_2,$ makes $F(x_1, x_2)$ in Eq.~(\ref{eq:mcF}) dominated by the sea parton
distributions, which are large at very small parton momenta. 

%%%%%%%%%%%%%%%%%%%%%%%%%%%%%%% fig. 3 %%%%%%%%%%%%%%%%%%%%

%fig.3: histogram + SSA and -SSA 20K 12<M<40 noi
\begin{figure}[ht]
\centering
\includegraphics[width=9cm]{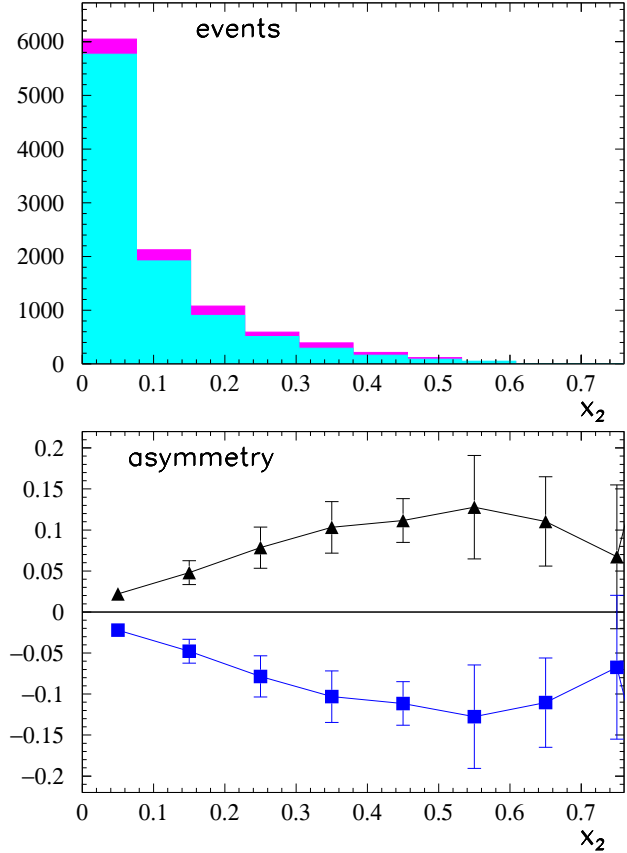}
\caption{The sample of $20\,000$ Drell-Yan events for the $pp^\uparrow \to \mu^+ 
\mu^- X$ reaction at $\sqrt{s}=200$ GeV, $12<M<40$ GeV, and $1<q_\sT <3$ GeV/$c$,
using the parametrization of Eq.~(\protect{\ref{eq:pTnoi}}) (see text). Upper
panel: for each bin in the parton momentum $x_2$ inside $p^\uparrow$, the darker 
histogram collects events with positive $\sin (\phi - \phi_{S_2})$ $(U)$, the 
superimposed lighter histogram collects the negative ones $(D)$. Lower panel: 
the asymmetry $(U-D)/(U+D)$; upward triangles for $N_u>0$ in 
Eq.~(\protect{\ref{eq:pTnoi}}) corresponds to a sign change in the Sivers function
from SIDIS to Drell-Yan processes; squares for $N_u<0$. Statistical error bars 
from 10 independent repetitions of the simulation. Continuous lines are drawn to 
guide the eye.}
\label{fig:noi12}
\end{figure}

%%%%%%%%%%%%%%%%%%%%%%%%%%%%%%%%%%%%%%%%%%%%%%%%%%%%%%%%%%%

%%%%%%%%%%%%%%%%%%%%%%%%%%%%%%%%%%%%%%%%%%%%%%%%%%%

\subsection{Single-spin asymmetries}
\label{sec:ssa}

In Fig.~\ref{fig:noi12}, the sample of $20\,000$ Drell-Yan events for the 
$pp^\uparrow \to \mu^+ \mu^- X$ reaction at $\sqrt{s}=200$ GeV is displayed for 
muon invariant mass in the $12<M<40$ GeV range. Results are reported in $x_2$ 
bins excluding the upper boundary $x_2 > 0.8$, which is scarcely or not at all 
populated, according to the upper band in Fig.~\ref{fig:scatter}. Events are 
accumulated according to Eq.~(\ref{eq:c4mc-noi}) based on the 
parametrization~(\ref{eq:pTnoi}) of the Sivers function; consequently, the 
transverse momentum distribution is constrained by $1<q_\sT <3$ GeV/$c$. In the 
upper panel, for each bin two groups of events are stored, one corresponding to 
positive values of $\sin (\phi - \phi_{S_2})$ in Eq.~(\ref{eq:mcS4}) (represented 
by the darker histogram), and one for negative values (superimposed lighter 
histogram). In the lower panel, the asymmetry $(U-D)/(U+D)$ is shown between the 
positive $(U)$ and negative $(D)$ values. Average values of the asymmetry and 
(statistical) error bars are obtained by 10 independent repetitions of the 
simulation. The upward triangles indicate the results assuming a positive 
normalization for the quark $u$ in Eq.~(\ref{eq:pTnoi}), which already takes into 
account the predicted sign change of $f_{1T}^\perp$ from Drell-Yan to 
SIDIS~\cite{Collins:2002kn} with respect to recent parametrizations of SIDIS
data~\cite{Anselmino:2005ea,Vogelsang:2005cs,Collins:2005wb}. For sake of 
comparison, the squares illustrate the opposite results that one would obtain 
ignoring such prediction. 

From the upper panel of Fig.~\ref{fig:noi12}, we deduce that the assumed 
elementary $\bar{q} q \to \g^\ast$ mechanism indeed populates the phase space for 
the lowest possible $\tau$, with more than 50\% of the events in the $0<x_2<0.1$ 
bin, leaving a $\sim 1/x_2$ distribution outside. In the lower panel, 
correspondingly, the error bars are small for $x_2<0.5$ and allow for a clean 
reconstruction of the asymmetry shape and, more importantly, for a conclusive test 
of the predicted sign change in $f_{1T}^\perp$. With the considered sample of 
$20\,000$ events, the same conclusion is not possible using the parametrization of 
Eq.~(\ref{eq:pTanselm}), because the asymmetry produced by 
Eq.~(\ref{eq:c4mc-anselm}) is too small. More quantitatively, half of the 
displayed error bar representes the variance $\Delta A(x_2)$ for the asymmetry 
$A=(U-D)/(U+D)$ in each $x_2$ bin. The results in the lower panel of 
Fig.~\ref{fig:noi12} can be approximated by the relation $\Delta A \approx 0.05\, 
x_2$. The asymmetry is statistically not compatible with zero if $A(x_2) > \Delta 
A=0.05\, x_2$ for the considered $x_2$ range. With $20\,000$ events, this condition 
is fulfilled only by the parametrization~(\ref{eq:pTnoi}), but not by the one in 
Eq.~(\ref{eq:pTanselm}). Finally, from Tab.~\ref{tab:rates} we deduce that an
hypothetical experiment in these kinematic conditions should run from one week to,
at most, almost one month in order to reach the indicated statistical error bars.

%%%%%%%%%%%%%%%%%%%%%%%%%%%%%%% fig.4 %%%%%%%%%%%%%%%%%%%%%%%%%
 
%fig.4: histogram + SSA and -SSA 100K 12<M<40 Anselmino

\begin{figure}[ht]
\centering
\includegraphics[width=8cm]{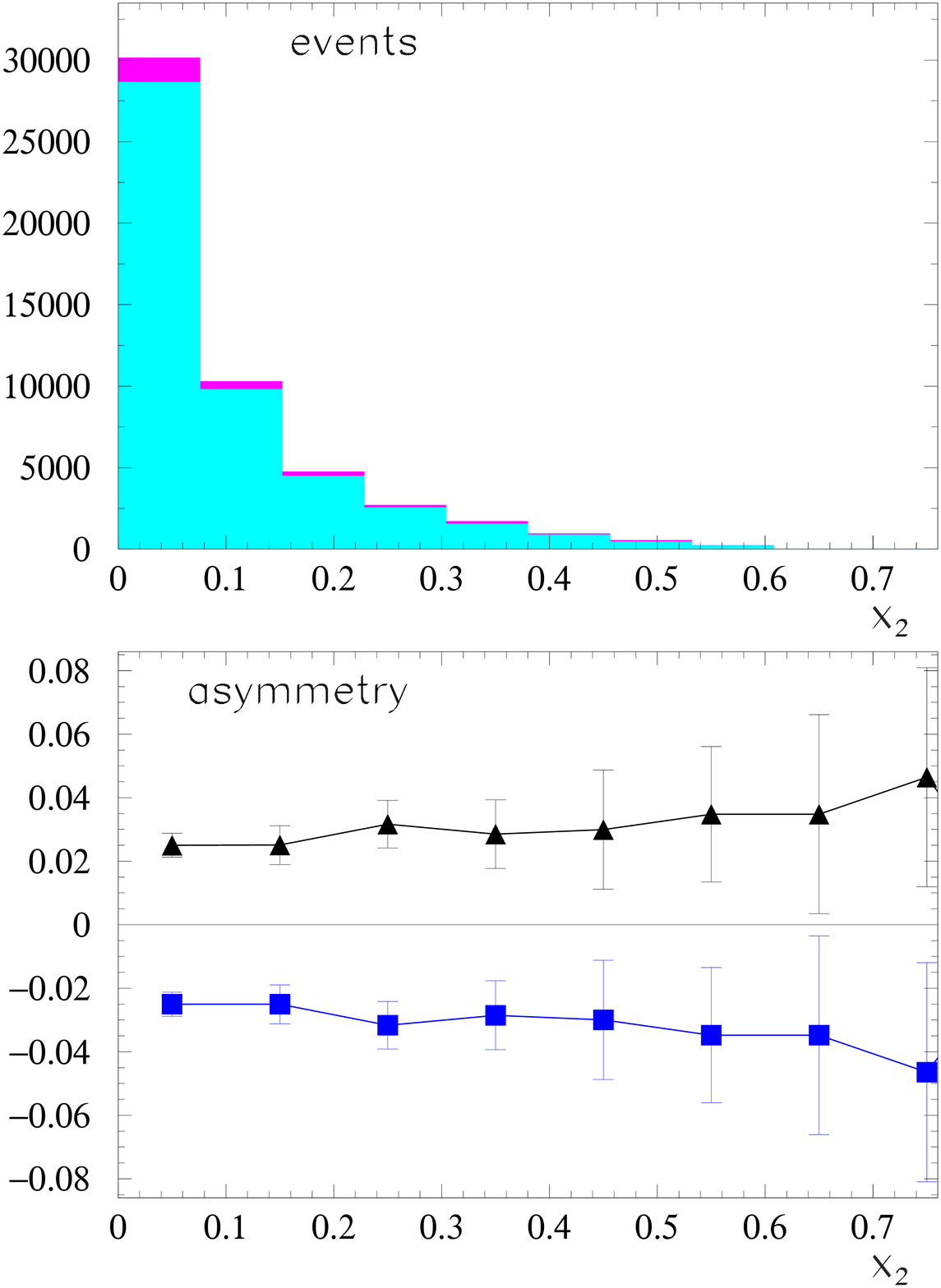}
\vspace{-.5cm}
\caption{Same as in Fig.~\protect{\ref{fig:noi12}} but considering $100\,000$ events 
using the parametrization of Eq.~(\protect{\ref{eq:pTanselm}}) and $0.1<q_\sT <2$ 
GeV/$c$ (see text). Lower panel: upward triangles for $N_u>0$ in 
Eq.~(\protect{\ref{eq:c4mc-anselm}}), squares for $N_u<0$.}
\label{fig:ans12}
\end{figure}

%%%%%%%%%%%%%%%%%%%%%%%%%%%%%%%%%%%%%%%%%%%%%%%%%%%%%%%%%%%%%%%
%%%%%%%%%%%%%%%%%%%%%%%%%%%%%%  fig. 5 %%%%%%%%%%%%%%%%%%%%%%%

%fig.5: histogram 100K 4<M<9 Anselmino + noi

\begin{figure}[h]
\centering
\includegraphics[width=9.5cm]{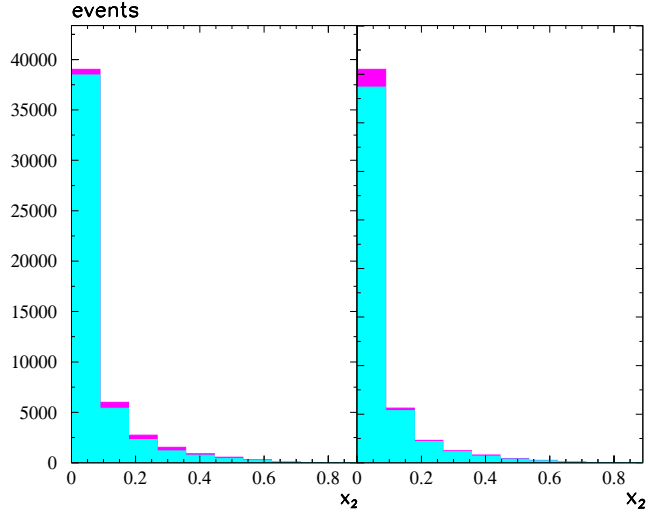}
\vspace{-.5cm}
\caption{The sample of $100\,000$ Drell-Yan events for the same reaction in the same
conditions and notations as in Fig.~\protect{\ref{fig:noi12}} but for $4<M<9$ GeV.
Left panel: parametrization of Eq.~(\protect{\ref{eq:pTnoi}}); right panel: 
parametrization of Eq.~(\protect{\ref{eq:pTanselm}}) (see text).}
\label{fig:histo4}
\end{figure}

%%%%%%%%%%%%%%%%%%%%%%%%%%%%%%%%%%%%%%%%%%%%%%%%%%%%%%%%%%%%%%

In Fig.~\ref{fig:ans12}, we consider the same kinematic conditions of the previous
Fig.~\ref{fig:noi12} but with a sample of $100\,000$ events, which would require at
least one month of running time, or four months in the most disfavoured conditions.
We employ the parametrization of Eq.~(\ref{eq:pTanselm}). We recall that the 
$q_\sT$ distribution is now integrated in the range $0.1<q_\sT <2$ GeV/$c$ with a 
resulting lower $\langle q_\sT^2 \rangle$ (see previous Sec.~\ref{sec:anselmino}). 
Notations in the figure are the same as in the previous Fig.~\ref{fig:noi12}. In
particular, in the lower panel upward triangles identify the results from
Eq.~(\ref{eq:c4mc-anselm}) with $N_u>0$ from Tab.~\ref{tab:pTanselm}; squared
points refer to the opposite choice. Therefore, we conclude that the 
parametrization~(\ref{eq:pTanselm}) demands for a much higher statistics in order 
to get a clear nonvanishing shape, because it produces an overall smaller 
asymmetry. Still, a definite answer is possible about the sign change 
prediction of $f_{1T}^\perp$ if the statistical sample meets the required 
conditions.

In Fig.~\ref{fig:histo4}, we show just the histogram of collected $100\,000$ events in
the same conditions and notations as before but for the lower $4<M<9$ GeV range. 
From Tab.~\ref{tab:rates}, we note that the necessary running time is $2\div 3$ 
times shorter than the one for collecting $20\,000$ events at the higher $12<M<40$ 
GeV range, depending on the parametrization chosen for $f_1$. Again, this is due 
to the already mentioned $1/\tau$ factor of the elementary $\bar{q} q \to \g^\ast$ 
mechanism, which privileges lower $\tau$. For the very same reason, an even larger 
portion of events (77\%) is contained in the first $0<x_2<0.1$ bin, while the
remaining 23\% is distributed for $x_2>0.1$ approximately as $1/x_2$. 

%%%%%%%%%%%%%%%%%%%%%%%%%%%%%% fig. 6 %%%%%%%%%%%%%%%%%%%%%%%%%

%fig 6: overlap corresponding SSA and -SSA

\begin{figure}[h]
\centering
\includegraphics[width=9cm]{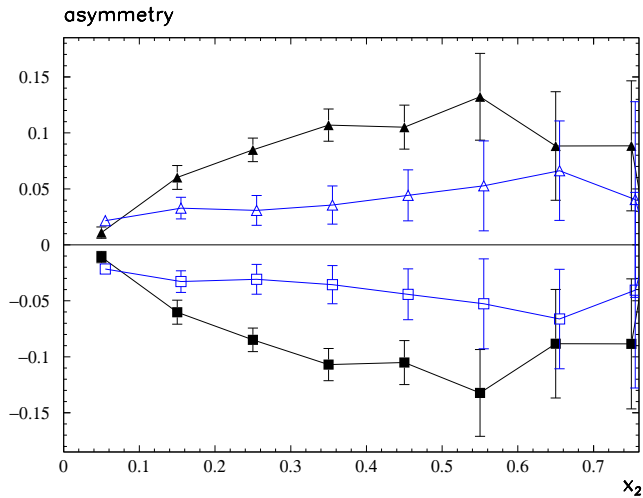}
\caption{The asymmetry $(U-D)/(U+D)$ corresponding to the histograms of
Fig.~\protect{\ref{fig:histo4}}, where $U$ identifies the darker histograms and $D$
the superimposed lighter ones (see text). Upward triangles for the 
parametrization of Eq.~(\protect{\ref{eq:pTnoi}}) with $N_u>0$; squares for
$N_u<0$. Open upward triangles for the parametrization of 
Eq.~(\protect{\ref{eq:pTanselm}}) with $N_u>0$; open squares for $N_u<0$.}
\label{fig:ssa4}
\end{figure}

%%%%%%%%%%%%%%%%%%%%%%%%%%%%%%%%%%%%%%%%%%%%%%%%%%%%%%%%%%%%%%

In Fig.~\ref{fig:ssa4}, we plot the spin asymmetries corresponding to the
histograms of the previous Fig.~\ref{fig:histo4}. Notations are the following.
Upward triangles correspond to the histograms in the left panel of
Fig.~\ref{fig:histo4}, i.e. to the parametrization of Eq.~(\ref{eq:pTnoi}); the
flavor dependent normalization is $N_u>0$, according to the properties of
$f_{1T}^\perp$ as it is extracted from SIDIS data. The corresponding choice $N_u<0$
is represented by the squares. The open upward triangles and squares show the
results for the other parametrization Eq.~(\ref{eq:pTanselm}). The accumulation of
events for very low $x_2$ values, which is evident in Fig.~\ref{fig:histo4}, here
reflects in very tiny error bars for the same bins, allowing to clearly distinguish
the two different parametrizations for $0<x_2\lesssim 0.5$. In this range, a
measurement of the Sivers effect for invariant masses as low as $4<M<9$ GeV, allows
to reduce the theoretical uncertainties in very few days of running time (from 2 to
12, depending on the parametrization of the unpolarized parton distributions).
Moreover, and most important, in the same $x_2$ range the statistical accuracy is
sufficient to directly test the predicted sign change of 
$f_{1T}^\perp$~\cite{Collins:2002kn}, irrespective of the uncertainty in the
theoretical input. 

For $0.5 \lesssim x_2$, the width of the error bars does not always allow for such
analysis, since the variance grows approximately as $\Delta A \approx 0.04 \, x_2$.
We already noted that for $12<M<40$ GeV approximately half of the $20\,000$ events lie
in the valence range $0.1<x_2 <0.7$ (see the upper panel in Fig.~\ref{fig:noi12}),
while for $4<M<9$ GeV $23\,000$ events out of $100\,000$ are found in the same range (see
Fig.~\ref{fig:histo4}). If we assume that the size of the error bars is
proportional to $1/\sqrt{N}$, with $N$ the number of events, then, in the
valence region for a given parametrization, the size of the error bars for the
higher $M$ range should approximately scale as $\sqrt{23/10} \approx 1.5$ with
respect to the one for the lower $M$ range. It is easy to check from the upward
triangles in Fig.~\ref{fig:noi12} and Fig.~\ref{fig:ssa4} that for the
parametrization~(\ref{eq:pTnoi}) indeed this approximate relation is verified. But 
from Tab.~\ref{tab:rates}, we deduce also that in the same running time necessary
to collect $20\,000$ events at $12<M<40$ GeV it is possible to collect $200\,000\div
300\,000$ events at $4<M<9$ GeV, depending on the chosen parametrization. In the
valence range, this means $46\,000\div 70\,000$ events (the 23\% of the total, as
before), which induces a reduction factor $2\div 2.5$ in the size of the error 
bars. Therefore, we can put a sort of "normalization" for the approximate 
behaviour of the size of the variance, by guessing that
\begin{align}
\Delta A(x_2) \approx 0.05 \, x_2 \, \sqrt{\frac{20\,000}{N}} &\mbox{\hspace{2cm}}
12<M<40 \; \mbox{GeV} & 0.1<x_2<0.7 \\
\Delta A(x_2) \approx 0.04 \, x_2 \, \sqrt{\frac{100\,000}{N}} &\mbox{\hspace{2cm}}
4<M<9 \; \mbox{GeV} & 0.1<x_2<0.7 \; .
\label{eq:variance}
\end{align}
For the lowest $0<x_2<0.1$ bin the coefficient is 0.008 and 0.004, respectively,
and the scale factor in the size of the error bars gets amplified.

%%%%%%%%%%%%%%%%%%%%%%%%%%%%%%%  fig. 7 %%%%%%%%%%%%%%%%%%%%%%

%fig.7: histogram+SSA for 100K 4<M<9 Anselmino and 0<x<02

\begin{figure}[h]
\centering
\includegraphics[width=9cm]{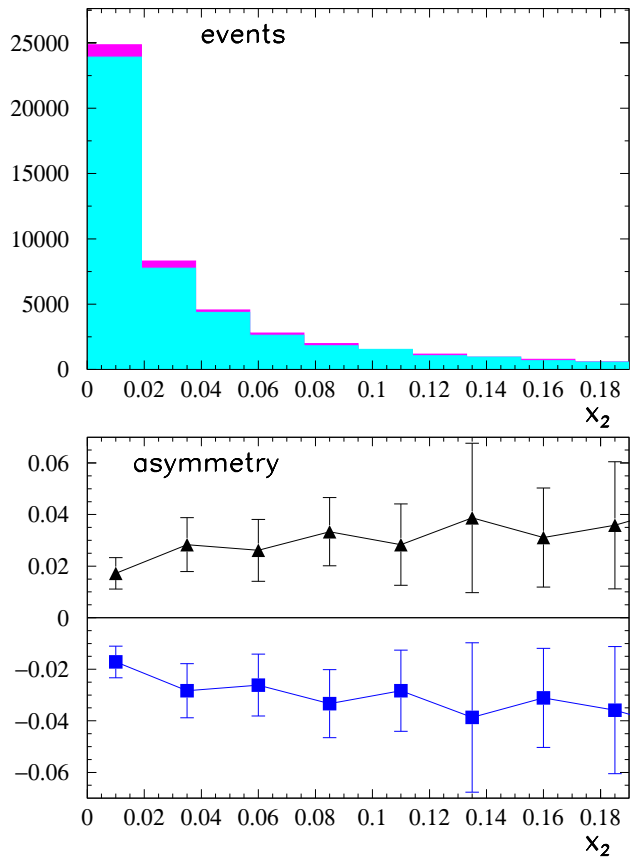}
\caption{The asymmetry corresponding to the open upward triangles and open
squares in Fig.~\protect{\ref{fig:ssa4}}, is shown in the $0<x_2<0.2$ range with a
finer binning.}
\label{fig:ans4sea}
\end{figure}

%%%%%%%%%%%%%%%%%%%%%%%%%%%%%%%%%%%%%%%%%%%%%%%%%%%%%%%%%%%%%%

In Fig.~\ref{fig:ans4sea}, we show a finer binning of the range
$0<x_2<0.2$ for the asymmetry obtained with the parametrization~(\ref{eq:pTanselm})
for $100\,000$ events in the $4<M<9$ GeV case. It is a closer view of the very tiny 
error bars of open upward triangles and squares in Fig.~\ref{fig:ssa4}. This is a
peculiar feature of this parametrization, which emphasizes the role of very low
$x_2$ through the parameters in Tab.~\ref{tab:pTanselm}, as it is discussed in
Sec.~\ref{sec:anselmino}.

%%%%%%%%%%%%%%%%%%%%%%%%%%%%%%%%%%%%%%%%%%%%%%%%%%%
\section{Conclusions}
\label{sec:end}

In this paper, we have concentrated on the investigation of the spin structure of
the proton using the single-polarized Drell-Yan process $pp^\uparrow \to \mu^+
\mu^- X$. At leading twist, the cross section contains several terms that lead to 
an asymmetric distribution of the final muon pair in its azimuthal angle $\phi$ 
with respect to the production plane. In previous 
papers~\cite{Bianconi:2004wu,Bianconi:2005bd}, we considered those terms involving 
the transversity distribution $h_1$ and the Boer-Mulders function
$h_1^\perp$~\cite{Boer:1999mm}, which is believed to be responsible for the very
well known violation of the Lam-Tung sum rule in the unpolarized Drell-Yan
data~\cite{Falciano:1986wk,Guanziroli:1987rp,Conway:1989fs}. There, we set up a
Monte Carlo to numerically simulate (polarized) antiproton-proton Drell-Yan 
collisions to study the best kinematic conditions for the HESR at 
GSI~\cite{Maggiora:2005cr,pax2} that allow to extract unambiguous information on 
the target parton distributions. 

Here, we have followed the same approach to isolate the contribution of the term
involving the Sivers function $f_{1T}^\perp$, a "naive T-odd" TMD partonic density 
that describes how the distribution of unpolarized quarks is distorted by the 
transverse polarization of the parent hadron. As such, $f_{1T}^\perp$ contains
unsuppressed information on the orbital motion of hidden confined partons; better, 
it contains information on their spatial distribution~\cite{Burkardt:2003je}, and 
it offers a natural link between microscopic properties of confined elementary 
constituents and hadronic measurable quantities, such as the nucleon anomalous 
magnetic moment~\cite{Burkardt:2005km}. Factorization theorems for TMD distribution
and fragmentation functions indicate that $f_{1T}^\perp$ is universal modulo a 
sign change (when switching from the SIDIS to the Drell-Yan process), due to a 
peculiar feature under the time-reversal operation of the gauge link operator
required to make its definition color-gauge invariant~\cite{Collins:2002kn}. 

Recently, very precise data for SSA involving $f_{1T}^\perp$ (the Sivers effect)
have been obtained for the SIDIS process on transversely polarized 
protons~\cite{Diefenthaler:2005gx}. This allowed for more realistic
parametrizations of 
$f_{1T}^\perp$~\cite{Anselmino:2005ea,Vogelsang:2005cs,Collins:2005wb}, that have 
been used then to make predictions for SSA in proton-proton collisions at RHIC 
(for a comparison among the various approaches, see also 
Ref.~\cite{Anselmino:2005an}). 

Here, we have numerically simulated the Sivers effect for the $pp^\uparrow \to
\mu^+ \mu^- X$ process at $\sqrt{s}=200$ GeV including the foreseen upgrade in the RHIC
luminosity (RHIC II). The goal is to explore the 
sensitivity of the simulated asymmetry to different input parametrizations, as 
well as to directly verify, within the reached statistical accuracy, the predicted 
sign change in the universality properties of the Sivers function. Therefore, we
have employed the parametrization of Ref.~\cite{Anselmino:2005ea} and a new high-energy 
parametrization of $f_{1T}^\perp$, whose flavor-dependent normalization
and $p_\sT$ distribution are constrained by recent RHIC data on SSA for the 
$pp^\uparrow \to \pi X$ process at $\sqrt{s}=200$ GeV~\cite{Adler:2005in}. The main 
difference is that the former, fitted to the SIDIS data of 
Ref.~\cite{Diefenthaler:2005gx}, displays an emphasized relative importance of the 
unfavoured $d$ quark, and it gives an average transverse momentum 
$\langle q_\sT \rangle$ of the lepton pair much lower than the latter. 
Consistently, we have built spin asymmetries by integrating the $q_\sT$ distribution
with adequate cutoffs, namely $0.1<q_\sT < 2$ GeV/$c$ for the former
parametrization, and $1<q_\sT<3$ GeV/$c$ for the latter. Results have been presented
as binned in the parton momenta $x_2$ of the polarized proton, i.e. by integrating
also upon the antiparton partner momenta $x_1$ and the zenithal muon pair 
distribution $\theta$ with no further cuts. 

Sorted events are divided in two groups, 
corresponding to opposite azimuthal orientations of the muon pair with respect to 
the reaction plane (conventionally indicated with $U$ and $D$), and the asymmetry
$(U-D)/(U+D)$ has been considered. Two different samples of $20\,000$ and $100\,000$
events have been selected, and statistical errors for $(U-D)/(U+D)$ have been 
obtained by making 10 independent repetitions of the simulation for each individual 
case and, then, calculating for each $x_2$ bin the average asymmetry and the 
variance. We have considered two different ranges of muon pair invariant 
mass, namely $4<M<9$ GeV and $12<M<40$ GeV. In this way, we avoid
overlaps with the resonance regions of the $\bar{c}c$ and $\bar{b}b$ quarkonium
systems, and we can safely assume that the elementary annihilation proceeds through
the $\bar{q}q\to \g^\ast$ mechanism. In particular, the Monte Carlo is based on the
corresponding leading-twist cross section. In the higher mass range, this 
theoretical analysis appears well established, since higher twists may be suppressed
as $M_p/M$, where $M_p$ is the proton mass. More questionable is the case of the
lower $M$ range, but this uncertainty can be traded for the much higher statistics
that can be reached because of the $1/M^2$ contribution of the $\g^\ast$ propagator.
Indeed, approximately 95\% of the events fall in the $4<M<9$ GeV range with a
significant reduction of the running time necessary to reach a predefined 
statistical accuracy. 

More generally, a very small portion of the phase space, corresponding to the lowest
possible values of $\tau = x_1 x_2 = M^2/s$, contains most of the events, also
because of the high cm square energy $s$; typically, $\langle x_{1/2} \rangle \sim 
0.01$. This is emphasized by the fact that in $pp$ collisions at least one of the 
two annihilating partons comes from the proton sea distribution, which is peaked at 
very low $x$ values. The importance of parton momenta below the valence region is 
potentially relevant to the theoretical models. In our case, it turns out that the
parametrization of Ref.~\cite{Anselmino:2005ea} gives small asymmetries because the
emphasized unfavoured $\bar{d} d$ annihilation is statistically suppressed. As a
consequence, in the $12<M<40$ GeV range $20\,000$ events are not sufficient to produce
a Sivers effect that is not statistically consistent with zero. On the contrary,
with the high-energy parametrization described in this paper this analysis is
possible in the range $0<x_2 \lesssim 0.5$, and a direct test of the "universal"
sign change of $f_{1T}^\perp$ can be unambiguously performed. 

A more favourable situation is encountered with lower $M$ values, say in the $4<M<9$
GeV range. The much higher statistics significantly reduces the error bars and
allows for a very short running time, typically as short as few days to collect the
$100\,000$ events simulated here at the foreseen luminosity of $10^{32}$ cm$^{-2}$ 
sec$^{-1}$, irrespectively of the choice of input parametrizations for 
$f_{1T}^\perp$. Moreover, we observe that in the $0<x_2\lesssim 0.5$ range the two 
choices give two clearly distinct asymmetries, and for each case the results with 
opposite signs in the Sivers function can be unambiguously separated. Remarkably, we
stress that this analysis holds also at very low $x_2$ values, typically as low as
0.01, which are relevant at RHIC. 

In conclusion, at the foreseen luminosity of $10^{32}$ cm$^{-2}$ sec$^{-1}$ (RHIC II) 
and with a dilution factor 0.5, in few days RHIC can collect $100\,000$ Drell-Yan 
events for the process $pp^\uparrow \to \mu^+ \mu^- X$ at $\sqrt{s}=200$ GeV and with 
muon pair invariant masses in the $4<M<9$ GeV range. From the measured Sivers effect, 
it should be possible to extract information on the $x$ structure of the Sivers
function in the $0<x\lesssim 0.5$ range, particularly also at very low $x\sim 0.01$, as 
well as to test its "universal" sign change predicted in Ref.~\cite{Collins:2002kn}. At 
higher $M$ values, like $12<M<40$ GeV, the situation is theoretically more favourable 
because the higher twists are suppressed as $1/M$. However, the lower density in the 
phase space makes the running time much longer. It is still possible to perform the 
previous analysis with $100\,000$ events and to come to the same conclusions, but at the 
price of taking data for some months. With a reduced sample, e.g. of $20\,000$ events, 
this time is shortened to few weeks, but our analysis was possible for only one of the 
chosen parametrizations, the other one giving results compatible with zero.

%%%%%%%%%%%%%%%%%%%%%%%%%%%%%%%%%%%%%%%%%%%%%%%%%%%

\begin{acknowledgments}

This work is part of the European Integrated Infrastructure Initiative in Hadron
Physics project under the contract number RII3-CT-2004-506078.

\end{acknowledgments}

%%%%%%%%%%%%%%%%%%%%%%%%%%%%%%%%%%%%%%%%%%%%%%%%%%%

\bibliographystyle{apsrev}
\bibliography{hadron.bib}

\end{document}